\documentstyle[12pt,epsfig]{article}
\let\section=\subsection  \let\subsection=\subsubsection

\def\be{\begin{equation}}
\def\ee{\end{equation}}
\def\bea{\begin{eqnarray}}
\def\eea{\end{eqnarray}}

\def\bR{\mbox{\boldmath $R$}}

\begin{document}

\begin{center}

{\large \bf Exotic baryon states in topological soliton models
}\\[5mm]
H. Walliser $^{*}$ and V.B. Kopeliovich $^{**}$  \\[5mm]
{\small \it 
$^{*}$ Fachbereich Physik, Universit\"at Siegen,  
D57068 Siegen, Germany\\
$^{**}$ Institute for Nuclear Research, Russian Academy of Sciences, Moscow,
117312 Russia }

\end{center}

\begin{abstract}\noindent
The novel observation of an exotic strangeness $S=+1$
baryon state at $1.54$ GeV will trigger an intensified search
for this and other baryons with exotic quantum numbers. 
This state was predicted long ago in topological soliton models.
We use this approach together with the new datum in order to
investigate its implications for the baryon spectrum. In
particular we estimate the positions of other pentaquark
and septuquark states with exotic and with non-exotic
quantum numbers.
\end{abstract}

\bigskip

\leftline{PACS: 12.39.Dc}

\leftline{Keywords: pentaquarks, septuquarks, anti-decuplet, solitons}

\vspace{1cm}

\section{Introduction}

In a quite recent paper \cite{japan}, Nakano et. al. report on 
an exotic strangeness $S\!=+1$ baryon state observed as a sharp resonance at
$1.54 \pm 0.01$GeV in photo-production from neutrons. The confirmation
of this finding would give formidable support to topological soliton models 
\cite{skyrme,witten} for a description of baryons in the 
non-perturbative regime of QCD. Higher multiplets containing states
carrying exotic quantum numbers arise naturally in the $SU(3)$
version of these models. 
These were called {\em exotic\/} because, within quark models,
such states cannot be built of only 3 valence quarks and additional 
quark-antiquark pairs must be added. So, the terms
{\em pentaquark} and {\em septuquark} characterize
the quark contents of these states.
Strictly, in soliton models there is nothing
exotic about these states, they just come as members of the next
higher multiplets. 

Indeed, beyond the minimal $\{8\}$ and $\{10\}$ baryons, also a 
$\{\overline{10}\}$ baryon multiplet was mentioned 
early by Chemtob \cite{chemtob}. Within a simple $SU(3)$ symmetric
Skyrme model Biedenharn and Dothan \cite{bieden} estimated
the excitation energy of the $\{\overline{10}\}$ with spin
$J=1/2$ to be only $0.60$ GeV (sic!) above the nucleon.
This  multiplet and a $\{27\}$ with spin $J=3/2$ both contain
low lying $S\!=+1$ states, called $Z$ and $Z$* in the following. 
First numbers for these exotic states taking the
configuration mixing caused by symmetry breaking into
account were given in
\cite{hans}, albeit some $0.1$ GeV too high if the value found in
\cite{japan} proves correct. Diakonov, Petrov and Polyakov \cite{dpp}
postulated the experimental
$P11(1.71)$ nucleon resonance a member
of the $\{\overline{10}\}$ multiplet and by this the
$Z$ again with low excitation energy ($0.59$ GeV). 
Weigel \cite{herbert} showed that similar low numbers ($0.63$ GeV) may
be obtained in an extended Skyrme model calculation, which 
includes a scalar field.

It should be added, that the excitation energies of 
similar exotic states have been 
estimated for arbitrary baryonic numbers \cite{VK}. 
It turned out, that all these states appear to be above 
threshold for the decay due 
to strong interactions. In general the excitation energies for the 
$B > 1$ systems are comparable to
those for baryons, e.g. the
$S=1$ dibaryon state belonging to the $\{\overline{35}\}$ multiplet 
was calculated, to be only $0.59$ GeV above $NN$-threshold
\cite{KSS}.

In this paper we address the following questions concerning
the $B=1$ sector.
Is an exotic $Z$ at $1.54$ GeV as reported in
\cite{japan} compatible with soliton models and the known
baryon spectrum? Provided the $Z$ is actually located
at this position, what does it imply for the other exotic states?

\section{$SU(3)$ soliton model}

There exists a large number of different soliton models, pure
pseudoscalar ones, models with scalar fields and/or
vector and axial-vector mesons and even models which include
quark degrees of freedom. There comes also a vast number of 
possible terms in the effective action for each of these models,
partly with free adjustable parameters. However, the $SU(3)$ symmetric
part always leads to the same collective hamiltonian with only 2 model
dependent quantities determining the baryon spectrum (section 2.1).
Unfortunately the situation for the symmetry breaking part is
less advantageous, but still there appears one dominating standard
symmetry breaker which will be the third model dependent quantity
needed (section 2.2). Thus, instead of refering to a specific model
(which comes with a number of free parameters as well) we are
going to adjust these 3 quantities to the known $\{8\}$ and
$\{10\}$ baryon spectra and to the just reported $Z$ \cite{japan}.
Using this input, we try to answer the questions posed in the
introduction. It will also be shown that the values needed for the
three quantities are not too far from what may be obtained in
a standard Skyrme model.

In the baryon sector, the static hedgehog soliton configuration
located in the non-strange $SU(2)$ subgroup 
is collectively and rigidly rotated in $SU(3)$ space. There
are other approaches like the soft rotator approach and the
bound state approach, but probably for $B=1$ the rigid rotator
approach is most appropriate.

\subsection{$SU(3)$ symmetric part}

The $SU(3)$ symmetric effective action
leads to the collective Lagrangian \cite{G}
\be 
L^S = - M + {1\over2}\Theta_{\pi}\sum_{a=1}^3 (\Omega_a^R)^2 +
{1\over2}\Theta_K \sum_{a=4}^7 (\Omega_a^R)^2 - 
{N_C B \over 2 \sqrt 3}\Omega_8^R \, . 
\ee
depending on the angular velocities $\Omega_a^R \, , a=1,\dots,8$. 
It is generic for all effective actions whose 
non-anomalous part contains at most two time derivatives, the term
linear in the angular velocity depends on the baryon number $B$
and the number of colors $N_C$ and it appears due to 
the Wess-Zumino-Witten anomaly. 

The soliton mass $M$, the pionic and kaonic moments of inertia 
$\Theta_{\pi}$ and $\Theta_K$ are model dependent quantities. 
The latter two are relevant to the baryon spectrum,
the soliton mass $M$, subject to large quantum corrections, 
enters the absolute masses only.
With the right and left angular momenta
\be
R_a = -{ \partial L^0 \over \partial \Omega_a^R}\;,\qquad \qquad 
L_a = \sum_{b=1}^8 D_{a b} R_b \;,
\ee
which transform according to Wigner functions $D_{a b}$
depending on the soliton's orientation,
the hamiltonian obtained by a Legendre transfomation 
\be\label{hs}
H^S = M + \frac{1}{2 \Theta_\pi} \bR^2
+ {1\over 2\Theta_K}\left(C_2(SU(3)) -
\bR^2 - N_C^2 B^2 / 12 \right) 
\ee
may be expressed by the second order Casimir operators of the $SU(3)$ group 
and its nonstrange $SU(2)$ subgroup
\be
C_2(SU(3))=\sum_{a=1}^8 R_a^2 \, , \qquad \bR^2=\sum_{a=1}^3 R_a^2 \, .
\ee
The eigenvalues of these operators for a given $SU(3)$ irrep $(p,q)$ with 
dimensionality $N=(p+1)(q+1)(p+q+2)/2$ are
\bea
C_2(SU(3)) |\{N\} (p,q),(Y_R J J_3) \rangle&=&  
\left[ {p^2+q^2+pq \over 3} +p+q \right]
|\{N\} (p,q),(Y_R J J_3)  \rangle \nonumber \\
\bR^2 |\{N\} (p,q), (Y_R J J_3) \rangle  
&=&  J(J+1) |\{N\} (p,q),(Y_R J J_3)  \rangle \, ,
\eea
where $(Y_RJJ_3)$ denote the right hypercharge and the baryon's spin.
The latter relation is due to the hedgehog ansatz which connects the spin
to the right isospin.
The states are still degenerate with respect to the left (flavor)
quantum numbers $(YTT_3)$ suppressed here. 
The constraint
$R_8 = N_C B / 2 \sqrt{3}$ 
fixes $Y_R=N_C B/3$ \cite{G}
and is written as triality condition \cite{bieden}
\be
Y_{max} = \frac{p+2q}{3} = B+m,
\ee
with $Y_{max}$ representing the maximal hypercharge of the $(p,q)$ multiplet.
Thus, baryons belong to irreps of $SU(3)/Z_3$.
With the octet being the lowest $B=1$ multiplet, the number
of colors must be $N_C=3$. It also follows
a spin-statistics-baryon number relation $(-1)^{2J+B}=1$,
which for $B=1$ allows for half-integer spins only \cite{bieden}.

From a quark model point of view, the integer $m$
must be interpreted as the number of additional
$q\bar{q}$ pairs present in the baryon state \cite{VK}. 
When $B=1$, we obtain for $m=0$ the minimal multiplets $\{8\}$ and $\{10\}$,
for $m=1$ the family of pentaquark
multiplets $\{\overline{10}\}$, $\{27\}$, $\{35\}$, $\{28\}$,
and for $m=2$ the septuquark multiplets 
$\{\overline{35}\}$, $\{64\},\; \{81\},\; \{80\}$ and $\{55\}$ (Fig. 1).
For the masses of the multiplets $\{8\}\, J=1/2$, $\{10\}\, J=3/2$, 
$\{\overline{10}\}\, J=1/2$, $\{27\}\, J=3/2$ and
$\{\overline{35}\}\, J=3/2$
simple relations
\bea
&& M_{\{10\}} - M_{\{8\}} = 3/ 2 \Theta_\pi, \nonumber \\
&& M_{\{\overline{10}\}} - M_{\{8\}} = 3/ 2 \Theta_K, \nonumber \\
&& M_{\{27\}} - M_{\{10\}} = 1/ \Theta_K, \\
&& M_{\{\overline{35}\}} - M_{\{10\}} = 15/ 4 \Theta_K \nonumber
\eea
hold. It is noticed that the mass difference of the minimal multiplets
depends on $\Theta_\pi$ only
\footnote {\tenrm It was shown for arbitrary $B$ \cite{VK} that coefficient of 
$1/2\Theta_K$ in $(3)$, $C_2(SU(3))-\bR^2 - 3B^2/4 = 3B/2$ for any minimal 
multiplet with $p+2q=3B$; $N_c=3$. },
whereas the mass differences between minimal 
and non-minimal multiplets
depend on $\Theta_K$ and $\Theta_\pi$. 
\begin{figure}[h]
\label{triality}
\begin{center}
\epsfig{figure=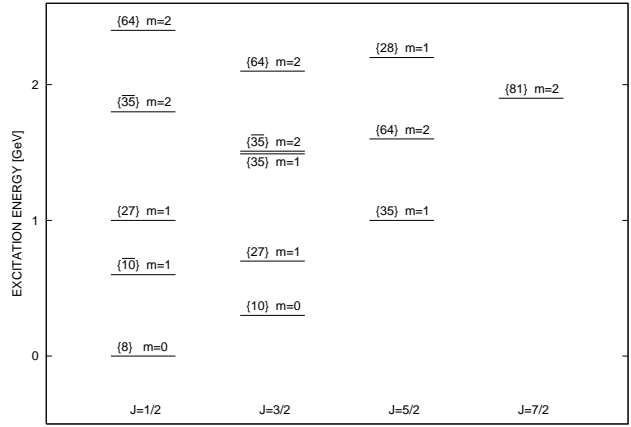,width=6cm,angle=270}
\protect\caption{$B=1$ baryon multiplets with less than $2.5$ GeV 
excitation energy for 
$\Theta_\pi = 5$ GeV$^{-1}$ and
$\Theta_K = 2.5$ GeV$^{-1}$. The number $m$ of additional
$q\bar{q}$ pairs is also given.
}
\end{center}
\end{figure}
With values 
$\Theta_\pi \simeq 5$ GeV$^{-1}$ and
$\Theta_K \simeq 2.5$ GeV$^{-1}$ from a naive 
Skyrme model the estimate 
$M_{\{\overline{10}\}} - M_{\{8\}} \simeq 0.60$ GeV \cite{bieden} 
was obtained, according to (7). 
The mass of the $\{27\}$ lies then $\simeq 0.10$ GeV higher.
In Fig. 1 we show
the spectrum of all baryon multiplets with an excitation energy up to
$2.5$ GeV using these moments of inertia for illustration.
The sequence of the lowest baryon multiplets
\be
\{ 8\}\,J=\frac{1}{2}\, , \enspace \{ 10\}\,J=\frac{3}{2}\, , \enspace
\{\overline{10}\}\,J=\frac{1}{2}\, , \enspace \{ 27\}\,J=\frac{3}{2}\, ,
\enspace \{ 35\}\,J=\frac{5}{2}\, \dots 
\ee
turns out to be unique within a large range of moments of inertia 
$\Theta_\pi /3 < \Theta_K < \Theta_\pi /2$, 
covering many realistic cases. Diagrams for the lowest non-minimal
baryon multiplets $\{\overline{10}\}$ and $\{ 27\}$ which
accomodate the interesting $S=+1$ states are depicted in 
Fig. 2.
%.........................................................
\begin{figure}[h]
\label{multiplet}
\setlength{\unitlength}{1.2cm}
\begin{flushleft}
\begin{picture}(12,6)
\put(3,3){\vector(1,0){2.5}}
\put(3,3){\vector(0,1){3}}
\put(2.6,5.7){$Y$}
\put(5.2,2.6){$T_3$}
\put(2,0){$\{\overline {10}\}\, J=1/2$}
\put(3.1,5.1){$Z^+$}

\put(3,5){\circle*{0.1}}
\put(2.5,4){\circle*{0.1}}
\put(3.5,4){\circle*{0.1}}
\put(2,3){\circle*{0.1}}
\put(3,3){\circle*{0.1}}
\put(4,3){\circle*{0.1}}
\put(1.5,2){\circle*{0.1}}
\put(2.5,2){\circle*{0.1}}
\put(3.5,2){\circle*{0.1}}
\put(4.5,2){\circle*{0.1}}

\put(1.5,2){\line(1,0){3}}
\put(1.5,2){\line(1,2){1.5}}
\put(4.5,2){\line(-1,2){1.5}}

\put(9,3){\vector(1,0){3}}
\put(9,3){\vector(0,1){3}}
\put(8.6,5.7){$Y$}
\put(11.7,2.6){$T_3$}
\put(8,0){$\{27\}\, J=3/2$}
\put(7.7,5.1){$Z$*$^0$}
\put(9.1,5.1){$Z$*$^+$}
\put(10.1,5.1){$Z$*$^{++}$}

\put(8,5){\circle*{0.1}}
\put(9,5){\circle*{0.1}}
\put(10,5){\circle*{0.1}}

\put(7.5,4){\circle*{0.1}}
\put(8.5,4){\circle*{0.1}}
\put(8.5,4){\circle {0.2}}
\put(9.5,4){\circle*{0.1}}
\put(9.5,4){\circle {0.2}}
\put(10.5,4){\circle*{0.1}}

\put(7,3){\circle*{0.1}}
\put(8,3){\circle*{0.1}}
\put(9,3){\circle*{0.1}}
\put(10,3){\circle*{0.1}}
\put(11,3){\circle*{0.1}}
\put(8,3){\circle {0.2}}
\put(9,3){\circle {0.2}}
\put(10,3){\circle {0.2}}
\put(9,3){\circle {0.3}}

\put(7.5,2){\circle*{0.1}}
\put(8.5,2){\circle*{0.1}}
\put(8.5,2){\circle {0.2}}
\put(9.5,2){\circle*{0.1}}
\put(9.5,2){\circle {0.2}}
\put(10.5,2){\circle*{0.1}}

\put(8,1){\circle*{0.1}}
\put(9,1){\circle*{0.1}}
\put(10,1){\circle*{0.1}}

\put(7,3){\line(1,2){1}}
\put(7,3){\line(1,-2){1}}
\put(8,5){\line(1,0){2}}
\put(8,1){\line(1,0){2}}
\put(11,3){\line(-1,2){1}}
\put(11,3){\line(-1,-2){1}}

\end{picture}

\caption{The $T_3-Y$ diagrams for the baryon multiplets 
$\{\overline{10}\}$ and $\{27\}$
which include the lowest $S=+1$ states.}
\end{flushleft}
\end{figure}
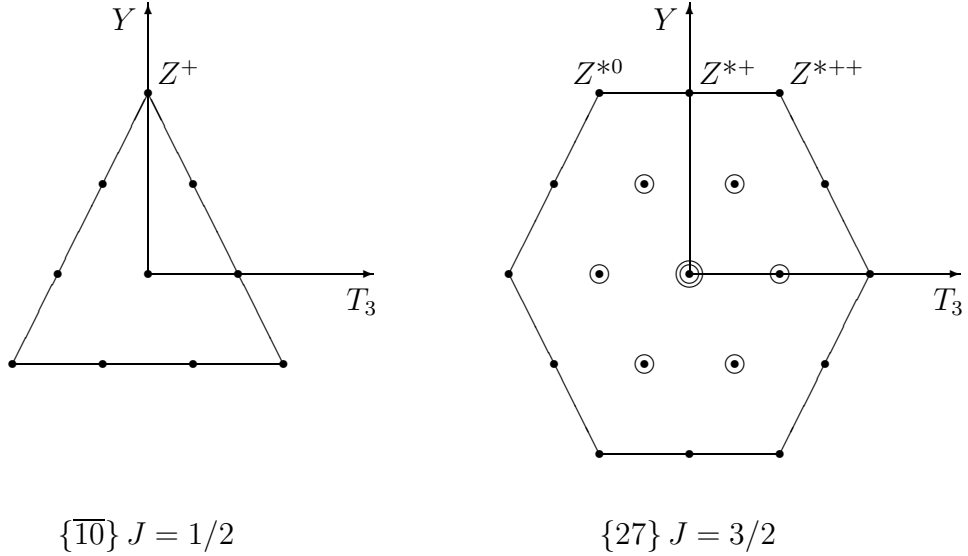
%.....................................................................

So far we have considered the $SU(3)$ symmetric case. In order to 
explain the splitting of baryon states within each multiplet we
have to take the explicit symmetry breaking into account.

\subsection{$SU(3)$ symmetry breaking}

The dominant standard symmetry breaker comes from mass and kinetic terms
in the effective action
which account for different meson masses and decay constants e.g.
$m_K \not= m_\pi$ and $F_K \not= F\pi$
\be
L^{SB} = -{1\over2} \Gamma (1-D_{88})
- \Delta \sum_{a=1}^3 D_{8a} \Omega_a^R + \dots
,
\ee
(first term).
There may be further terms of minor importance which depend on the 
specific effective action used. As an example we will optionally
include such a term which arises from $\rho-\omega$ mixing in vector meson
lagrangians (second term). This may serve as a test for the model
dependence of our results.
The corresponding hamiltonian is  
\be\label{hsb}
H^{SB} = {1\over2} \Gamma (1-D_{88}) - 
{\Delta \over \Theta_\pi}
 \sum_{a=1}^3 D_{8a} R_a + \dots \, .
\ee
The quantities $\Gamma$ and $\Delta$ are again model dependent
quantities, they determine the strength of symmetry breaking.
To begin with we consider only the standard symmetry breaker $\Gamma$.

It was early noticed that a perturbative treatment of this
symmetry breaker leads to a splitting 
$(M_\Lambda-M_N):(M_\Sigma-M_\Lambda):(M_\Xi-M_\Sigma)=2:2:1$
for the $\{8\}$ baryons \cite{chemtob,G} in variance with observation.
Because symmetry breaking is strong, eq. (\ref{hsb})
must be diagonalised
in the basis of the unperturbed eigenstates of $H^S$. 
By this procedure the states of a certain multiplet 
pick up components of higher representations. 
Nevertheless we will address also the mixed states
as $\{ 8\}$ states, $\{ 10\}$ states and so on, according to their
dominant contribution.

The best values for the moments of inertia $\Theta_\pi$ and $\Theta_K$
and the symmetry breaker $\Gamma$ are listed in Table 1 (fit A).
Optionally the symmetry breaker $\Delta$ is also included (fit B).
\begin{table}[h]
\label{fit}
\begin{center}
\caption{Moments of inertia and symmetry breakers as obtained
from a fit to the baryon spectrum includung the novel $Z$ datum.
}

\vspace{6mm}

\begin{tabular}{c|c|c|c|c|}
& $\Theta_\pi$[GeV$^{-1}$] & $\Theta_K$[GeV$^{-1}$]& 
$\Gamma$[GeV] & $\quad \Delta \quad $\\ [0.5ex]
\hline
&&&&\\
$\quad$ fit A $\quad$ & $5.61$ & $2.84$ & $1.45$ &  $-$ \\[0.5ex]
$\quad$ fit B $\quad$ & $5.87$ & $2.74$ & $1.34$ & $0.40$ \\[0.5ex]
\hline
\end{tabular}
\end{center}
\end{table}
First we show in Fig. 3 the dependence of the 
$Z$ and $Z$* energies on
the kaonic moment of inertia $\Theta_K$ with the other parameters
kept fixed. 
The sensitive dependence expected from eq. (7) persists when
symmetry breaking is included. If the experimental datum for $Z$ 
proves correct, a relatively large kaonic moment of inertia (Table 1) is
required.
\begin{figure}[h]
\label{moment}
\begin{center}
\epsfig{figure=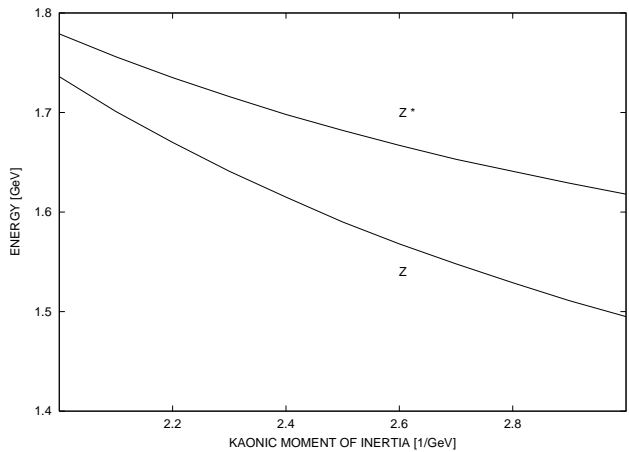,width=6cm,angle=270}
\protect\caption{The masses of the $S=1$ baryon masses $Z$ and
$Z$* depending on the kaonic moment of inertia. 
$\Theta_\pi=5.87$ GeV$^{-1}$ and $\Gamma=1.34$ GeV are kept fixed.
}
\end{center}
\end{figure}

Let us compare with what is obtained from a standard Skyrme model 
\cite{skyrme, witten} (the only parameter of this model $e=4.05$) 
with mass and kinetic symmetry breakers included with
mesonic parameters. There appear time derivatives in the kinetic
symmetry breaker which were neglected in \cite{hans}
(adiabatic approximation) with the argument
that they are suppressed by two orders in an $1/N_C$ expansion
and there should come many other symmetry breaking terms at this
order which are also not taken into account.
This leads to $\Theta_\pi=5.88$ GeV$^{-1}$, $\Gamma=1.32$ GeV
and a relatively small kaonic moment of inertia $\Theta_K=2.19$ GeV$^{-1}$
(connected with larger $Z$ and $Z$* masses, Fig. 3).
However, the non-adiabatic terms in the kinetic symmetry breaker
are not really small, giving a sizeable
contribution to the kaonic moment of inertia 
$\Theta_K=2.80$ GeV$^{-1}$ together with symmetry breaking terms and even
terms non-diagonal in the angular momenta. Since the latter were
never properly treated, these numbers should be compared 
with reservation to those given in Table 1.
Nevertheless, it seems that
the standard Skyrme model potentially may provide values close to fit B.
Relative to fit A, the standard symmetry breaker from the Skyrme model
appears too weak indicating that an important symmetry breaking piece
is missing in this model. Concluding this discussion, it should be
stressed, that the non-adiabatic terms in the kinetic 
symmetry breaker are of course not the only possibility to arrive at larger
kaonic moments of inertia. The inclusion of other degrees of freedom
or the consideration of additional terms in the effective action
sensitively influences this quantity. In this respect the position 
of the exotic $Z$ baryon proves an important constraint on soliton models.

The resulting baryon spectrum is shown in Fig. 4. 
It is noticed that
for fit A, with the standard symmetry breaker alone, (i) the $\Sigma-\Lambda$
mass difference is too large, (ii) the splitting in the $J=1/2$
multiplets relative to that in the $J=3/2$ multiplets is overestimated,
and (iii) the corresponding $SU(2)$ symmetry breaker
may account only for half the neutron-proton split (not shown here, see
e.g.\cite{hans}).
Essentially all three deficiencies may be cured by including the second
symmetry breaker, fit B. This does of course not mean, that the
additional symmetry breaker must be exacly of the form (\ref{hsb}), other
operator structures are possible. As mentioned, we include fit B
mainly to get a notion of the model dependence of our results.
\begin{figure}[h]
\label{spectrum}
\begin{center}
\epsfig{figure=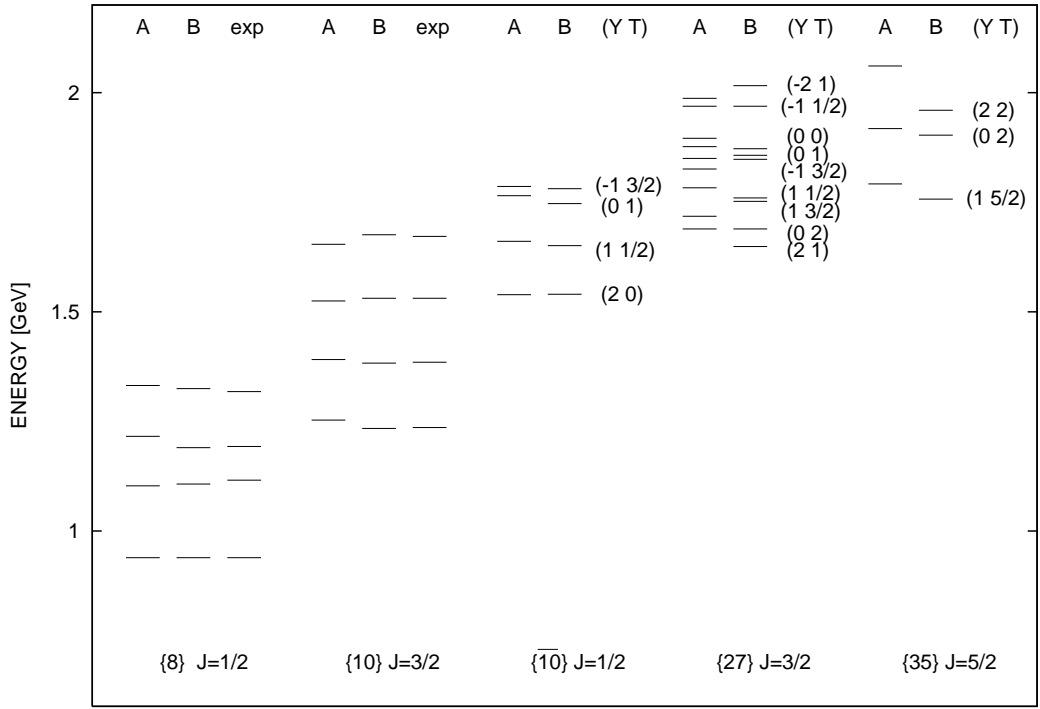,width=10cm,angle=270}
\protect\caption{Lowest rotational states in the $SU(3)$ 
soliton model for fits A and B. The experimental masses
of the $\{8\}$ and $\{10\}$ baryons are depicted for comparison.
Not all states of the $\{35\}$ are shown.
}

\end{center}
\end{figure}
It seems that the levels of the $\{\overline{10}\}$ are 
relatively stable in contrast to the
$\{ 27\}$ whose states depend sensitively on
the specific form of the symmetry breakers such that even the
ordering of the levels gets changed.

The lowest states of the $\{\overline{10}\}$ and $\{ 27\}$
are listed in Tables 2 and 3.
We distinguish states with exotic quantum numbers from
those with
non-exotic quantum numbers
$-2 \leq Y \leq 1$ and $T \leq 1 + Y/2$. Generally,
the former are "cleaner", because they cannot mix
with vibrational excitations (apart from their own radial
excitations). 
\begin{table}[h]
\label{exotic}
\begin{center}
\caption{Rotational states of non-minimal multiplets with exotic 
quantum numbers below $2$ GeV including all members of
$\{\overline{10}\}$ and $\{ 27\}$.
The experimental $Z$ datum enters the fits.
The lowest exotic $Y=\pm 3$
baryon states are also included.
}

\vspace{6mm}
 
\begin{tabular}{r|crcl|cc|}
& $\, J \,$ & $\, Y \, $ & $\, T \, $ & 
$\quad \,$ decay modes & \multicolumn{2}{c|}{estimated energy [GeV]} \\
&&&&& $\qquad$ A & B \\
\hline
$Z \quad \{\overline{10}\}\, $ & $\frac{1}{2}$ & $2$ & $0$ & 
$\qquad KN$ & $\qquad 1.54$ & $1.54$ \\[0.5ex] 
$Z$* $\, \{27\} \, $ & $\frac{3}{2}$ & $2$ & $1$ & 
$\qquad KN$ & $\qquad 1.69$ & $1.65$ \\[0.5ex] 
$\{27\}\,$ & $\frac{3}{2}$ & $0$ & $2$ & $\qquad \pi\Sigma$, 
$\pi\Sigma$*, $\pi\pi\Lambda$ & $\qquad 1.72$ & $1.69$ \\[0.5ex] 
$X \quad \{35\}\,$ & $\frac{5}{2}$ & $1$ & $\frac{5}{2}$ &$\qquad \pi\Delta$,
$\pi\pi N$ & $\qquad 1.79$ & $1.76$ \\[0.5ex] 
$\{\overline{10}\}\,$ & $\frac{1}{2}$ & $-1$ & $\frac{3}{2}$ & 
$\qquad \pi\Xi$, $\pi\Xi$*, $\bar K \Sigma \quad$  & 
$\qquad 1.79$ & $1.78$ \\[0.5ex]
$\{27\}\,$ & $\frac{3}{2}$ & $-1$ & $\frac{3}{2}$ & 
$\qquad \pi\Xi$, $\pi\Xi$*, $\bar K \Sigma \quad$  & 
$\qquad 1.85$ & $1.85$ \\[0.5ex]
$\{35\}\,$ & $\frac{5}{2}$ & $0$ & $2$ & $\qquad \pi\Sigma$, 
$\pi\Sigma$* & $\qquad 1.92$ & $1.90$ \\[0.5ex] 
$\{35\}\,$ & $\frac{5}{2}$ & $2$ & $2$ & $\qquad K\Delta$, $K\pi N$
& $\qquad 2.06$ & $1.96$ \\[0.5ex]
$\{27\}\,$ & $\frac{3}{2}$ & $-2$ & $1$ & $\qquad \pi\Omega$, 
$\bar K \Xi$, $\bar K \Xi$* & $\qquad 1.99$ & $2.02$ \\[0.5ex]  
\hline    
$\{35\}\,$ & $\frac{5}{2}$ & $-3$ & $\frac{1}{2}$ & 
$\qquad \bar K \Omega$, $\bar K \bar K \Xi$& $\qquad 2.31 $ & $2.36 $\\[0.5ex]    
$Z$** $\{\overline{35}\}\,$ & $\frac{3}{2}$ & $3$ & $\frac{1}{2}$ & 
$\qquad KKN$, $KK\Delta$  & $\qquad 2.41 $ & $2.38 $\\[0.5ex]    
\hline
\end{tabular}

\end{center}

\end{table}
Because additional vibrations 
on top of these states can only
enhance the energy, these turn out to be really the lowest states
with exotic quantum numbers starting with the $S\!=1$ baryon
states $Z$ and $Z$*. The latter are
experimentally accessible via the reactions
\bea
\gamma N \quad \longrightarrow &\bar K Z&
\longrightarrow \quad \bar K K N \nonumber \\
\pi N \quad \longrightarrow &\bar K Z&
\longrightarrow \quad \bar K K N \nonumber \\
N N \quad \longrightarrow &(\Lambda,\Sigma) Z& 
\longrightarrow  \quad (\Lambda,\Sigma) K N  \nonumber 
\eea
and in $KN$ scattering. The novel measurement \cite{japan} 
was a photo-production experiment of the first type.
The $S \not= 1$ exotics
are more difficult to measure, e.g. the
$X$ of Table 2 via the reactions
\bea
\pi N \quad \longrightarrow &\pi X&
\longrightarrow \quad \pi \pi \Delta \nonumber \\
N N \quad \longrightarrow &\Delta X&
\longrightarrow \quad \pi \Delta \Delta \, . \nonumber 
\eea
\begin{table}[h]
\label{normal}
\begin{center}
\caption{Rotational states of higher multiplets with
non-exotic quantum numbers below $2$ GeV including all members of the
$\{\overline{10}\}$ and $\{ 27\}$.
}

\vspace{6mm}

\begin{tabular}{r|crcl|cc|}
& $\, J \,$ & $\, Y \, $ & $\, T \, $ & 
$\qquad $ candidate & \multicolumn{2}{c|}{estimated energy [GeV]}\\
&&&&& $\qquad$ A & B \\
\hline
$N$* $\,\{\overline{10}\} $ & $\frac{1}{2}$ & $1$ & $\frac{1}{2}$ & 
$\qquad N \, P11(1.71)*** $ & $\qquad 1.66$ & $1.65$ \\ [0.5ex]
$\Sigma$* $\,   \{\overline{10}\} $ & $\frac{1}{2}$ & $0$ & $1$ & 
$\qquad \Sigma \, P11(1.77)* $  & $\qquad 1.77$ & $1.75$ \\[0.5ex]
$\Delta$*  $\, \{27\} \, $ & $\frac{3}{2}$ & $1$ & 
$\frac{3}{2}$ & & $\qquad 1.83$ & $1.75$ \\[0.5ex]
$\{27\}$ & $\frac{3}{2}$ & $1$ & $\frac{1}{2}$ & 
$\qquad N \, P13(1.72)**** $ & $\qquad 1.78$ & $1.76$ \\[0.5ex]
 $\{27\}$ & $\frac{3}{2}$ & $0$ & $1$ &
$\qquad \Sigma \, P13(1.84)* $ & $\qquad 1.90$ & $1.86$ \\[0.5ex] 
$\Lambda$* $\,  \{27\}$ & $\frac{3}{2}$ & $0$ & $0$ &
$\qquad \Lambda \, P03(1.89)**** $ & $\qquad 1.88$ & $1.87$ \\[0.5ex] 
$\{27\}$ & $\frac{3}{2}$ & $-1$ & $\frac{1}{2}$ & 
$\qquad \Xi \, ?? \, (1.95)*** $ & $\qquad 1.97$ & $1.97$ \\[0.5ex]
\hline
\end{tabular}

\end{center}

\end{table}
We included also the lowest exotic states with strangeness $S=+2$
and $S=-4$ with main components in the $\{\overline{35}\}$ and $\{35\}$
multiplets respectively. The $S=+2$ state $Z$** still can be produced in 
binary reactions, e.g. $ K^0 p \longrightarrow K^- Z$**$^{,++}$, but the 
energy of this state is already quite considerable, $\simeq 2.4$ GeV.
On the other hand, the $S=-4$ state is more difficult to produce,
but detection seems to be simpler because final $\Omega^-$ and $K^-$
are easy to see.

In contrast, the states with non-exotic quantum numbers 
in Table 3 mix strongly with vibrational
excitations of the $\{8\}$ and $\{10\}$ baryons. For example the $N$* 
rotational state, identified
with the nucleon resonance $P11(1.71)$ in \cite{dpp},
mixes strongly with a $2\hbar \omega$ radial excitation 
which may even lead to a doubling of states as found in
\cite{herbert}. This situation renders an easy interpretation
difficult. 
Probably the cleanest of these states with non-exotic quantum numbers
is the one called
$\Lambda$* which predominantly couples to
the {\em non-resonant\/} magnetic dipole mode. 
But even here it is not excluded
that the good agreement with the position of the experimental $\Lambda$ 
resonance $P03(1.89)$ is accidential. 
Also, there is not even a candidate 
for the rotational state called $\Delta$* 
listed by the PDG in the required energy region with the empirical
$\Delta$ resonance $P33(1.92)$ lying $\simeq 0.1$ GeV too high.
On the other hand in 5 cases we do have candidates close
to the estimated energies.
There is certainly some 
evidence that the numbers presented are not unreasonable.

It should be added that the energies for the $\{\overline{10}\}$
baryons presented here
differ substantially from what was obtained in ref.\cite{dpp}
using simple perturbation theory. Their $\{\overline{10}\}$
splitting is overestimated by more than a factor of 1.5 .

\section{The $S\!=1$ baryon spectrum}

So far we have considered rotational states only. The real situation
is complicated by the fact that there is a whole tower of vibrational
excitations connected with each of these rotational states. We will
briefly address this issue on a quite qualitative level
particularly for the $S\!=1$ sector. Possibly this may be of help
for experimentalists in search for further exotic baryons.

The lowest states in the $S\!=1$ sector are the rotational states
$Z$ and $Z$* discussed in the previous section. 
As mentioned, we believe that the energies of these 2 states 
should be close to each other with that of $Z$* somewhat 
larger ($\simeq 0.10-0.15$ GeV). Such rotational states appear
as sharp resonances with small widths relative to the broader
vibrational states. The width of $Z$ was given in \cite{japan} 
to be smaller than $25$ MeV, and that of $Z$* should be somewhat
larger due to phase space arguments. Probably the $Z$* will
be the next exotic state detected.

Certainly, in soliton models there exist radial excitations 
(breathing modes) for each rotational state.
For most of the $\{8\}$ and $\{10\}$ baryons such excitations 
correspond to wellknown resonances as e.g. the Roper resonance
for the nucleon. A breathing mode excitation energy  
$\simeq 0.45$ GeV for the $Z$ was calculated in \cite{herbert},
and that of $Z$* should be considerably smaller because the 
latter object is more extended
due to centrifugal forces connected to a larger spin (similar
situation as for Roper and the $\Delta$ resonance $P33(1.60)$).
Therefore we
may expect excited $P01$ and $P13$ states close together as indicated
in Fig. 5 (the order may be reversed!). 
\begin{figure}[h]
%\label{KN}
\begin{center}
\epsfig{figure=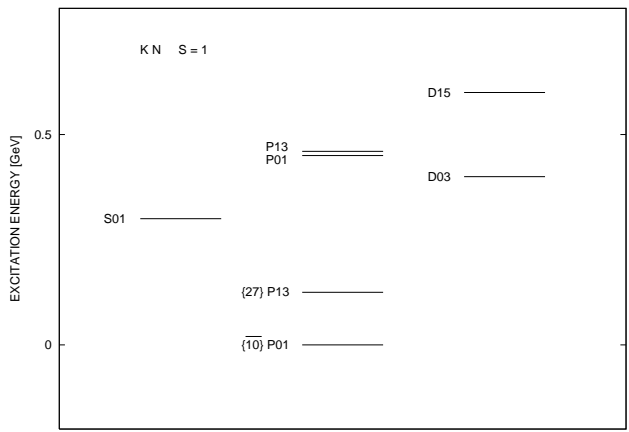,width=6cm,angle=270}
\protect\caption{Tentative baryon spectrum for the $S\!=1$ sector. 
}

\end{center}
\end{figure}
In addition,
there will be strong quadrupole excitations as those obtained
in soliton models \cite{bernd} and
seen empirically in the well studied $S\!=0$ and $S\!=-1$ sectors
(with roughly $0.4$ and $0.6$ GeV excitation energy).
In these sectors there appear also a number of S-wave resonances
through $\bar KN$, $K\Lambda$, $K\Sigma$ and $K\Xi$ bound states just
below the corresponding thresholds \cite{bernd}. 
Although such an interpretation
seems less clear in the $S\!=1$ sector, a
low lying $S01$ resonance is nevertheless expected, just
by inspection of the other sectors.

Tentatively, this leads to a $S\!=+1$ baryon spectrum depicted in
Fig.5. The T-matrix poles $P01(1.83)$, $P13(1.81)$, $D03(1.79)$ and
$D15(2.07)$ extracted from early $KN$ scattering experiments
\cite{said} qualitatively would fit with such a scheme, the spacings
however are considerably smaller than in Fig.5. So, in case 
these T-matrix poles prove correct,
a strong quenching of the spectrum shown in Fig. 5 has to be
understood. The existence of such poles, particularly in the
$D$-waves, would likewise favour a $Z$ located considerably below
these resonances compatible with the datum $1.54$ GeV.

\section{Conclusion}

We have shown that a low position of the exotic $S=+1$ baryon 
$Z$ with quantum numbers $J=1/2$ and $T=0$ at the reported
$1.54$ GeV is compatible with soliton models and the known
baryon spectrum. For all members of the 
$\{\overline{10}\}$ and $\{27\}$ multiplets with non-exotic
quantum numbers we find candidates close to the estimated
energies, with one exception: the empirical $\Delta$ resonance
$P33(1.92)$ lies $\simeq 0.1$ GeV too high. There will 
be a strong mixing of these states with vibrational modes of the 
$\{8\}$ and $\{10\}$ baryons, which may lead to considerable energy shifts
and even to a doubling of states. Also the T-matrix poles of
early $KN$ scattering experiments favour a low $Z$ sufficiently below
these resonances, with the caveat, that when these poles are correct
a strong quenching of the $S=1$ baryon spectrum compared to other
sectors has to be explained.

However, the soliton model by itself does not exclude a $Z$ baryon at 
higher energies.
Therefore the confirmation of this datum, which proves a stringent
constraint on these models, is most important.

Under the assumption, that the exotic $Z$ is actually located
at the reported position, we have estimated the energies of other
exotic baryons. First of all, there will be a
further $S=+1$ baryon $Z$* with quantum numbers $J=3/2$ and 
$T=1$, some $0.10-0.15$ GeV
above the $Z$. This will probably be the next state to be discovered
in similar experiments also as a sharp resonance with a somewhat
larger width yet. Moreover there will be a tower of vibrational
excitations built on these two exotic states, which should appear
as broader resonances several $0.1$ GeV above these energies.

There are also several low lying $S\not=1$ baryons
with exotic isospin, starting with a $J=1/2$ state with 
quantum numbers $S=0$ and $T=2$ at $\simeq 1.7$ GeV.
These states are more difficult to access experimentally.
The lowest $S=+2$ and $S=-4$ baryon states may also be of some interest
although they are already expected at high energies $\simeq 2.3-2.4$ GeV.

\vspace{5mm}

VBK is indebted to B.O.Kerbikov, A.E.Kudryavtsev, L.B.Okun' and other 
participants of the ITEP Thursday seminar for stimulating
discussions; his work was supported by RFBR, grant 01-02-16615.


\bigskip
\noindent

\begin{thebibliography}{80}

\bibitem{japan}    T. Nakano et. al., preprint, hep-ex/0301020. 
% exotic S=1                   


\bibitem{skyrme}   T.H.R. Skyrme,
                   Proc. Roy. Soc. A {\bf 260}, 127 (1961);
                   Nucl. Phys. {\bf 31}, 556 (1962).
% skyrmions


\bibitem{witten}   E. Witten, 
                   Nucl. Phys. B {\bf 223}, 422, 433 (1983).
% laege N_C 


\bibitem{chemtob}  M. Chemtob,
                   Nucl. Phys. B {\bf 256}, 600 (1985).
% {anti10} first mentioned


\bibitem{bieden}   L.C. Biedenharn and Y. Dothan, From $SU(3)$ to gravity
                   (Ne'eman Festschrift) (Cambridge Univ. Press 1986).
% triality, Z estimate


\bibitem{hans}     H. Walliser,
                   Nucl. Phys. A {\bf 548}, 649 (1992); and
                   in Baryons as Skyrme Solitons, p. 247,
                   ed. G. Holzwarth, World Scientific (1992).
% first config mix result


\bibitem{dpp}      D. Diakonov, V. Petrov, and M. Polyakov,
                   Z. Phys. A {\bf 359}, 305 (1997).
% {anti10} adjusted to P11(1710)


\bibitem{herbert}  H. Weigel,
                   Eur. Phys. J. A {\bf 2}, 391 (1998).
% scalar meson, radial excitations


\bibitem{VK}        V.B. Kopeliovich, 
                    Phys. Lett. B {\bf 259}, 234 (1991)
% rotation energies for arbitr. B


\bibitem{KSS}       V.B. Kopeliovich, B. Schwesinger, and B.E. Stern,
                    Nucl. Phys. A {\bf 549}, 485 (1992)
% dibaryon in 35              

\bibitem{G}        E. Guadagnini, Nucl. Phys. B236, 35 (1984).
% Y_R = B


\bibitem{bernd}    B. Schwesinger,
                   Nucl. Phys. A {\bf 537}, 253 (1992).
% vibrational modes SU(3)


\bibitem{said}     J.S. Hyslop, R.A. Arndt, L.D. Roper,
                   and R.L. Workman,
                   Phys. Rev. D {\bf 46}, 961 (1992).
% T-matrix poles

\end{thebibliography}
\end{document}